\documentclass[conference]{IEEEtran}
\usepackage{latexsym}
\usepackage{amsfonts}
\usepackage{amssymb}
\usepackage{amsbsy}
\usepackage{amsmath}
\usepackage[dvips]{epsfig}
\usepackage{pgf}
\usepackage{graphicx,subfigure}
\usepackage{balance}

\balance

\title{Network Coded Rate Scheduling\\ for Two-way Relay Networks}

\author{\IEEEauthorblockN{Dong Min Kim and Seong-Lyun Kim\\}
\IEEEauthorblockA{School of Electrical and Electronic Engineering, Yonsei University\\
50 Yonsei-ro, Seodaemun-gu, Seoul 120-749, Korea\\
Email: \{dmkim, slkim\}@ramo.yonsei.ac.kr} }

\begin{document}
\maketitle

\begin{abstract}
We investigate a scheduling scheme incorporating network coding and channel varying information for the two-way relay networks. Our scheduler aims at minimizing the time span needed to send all the data of each source of the network. We consider three channel models, time invariant channels, time varying channels with finite states and time varying Rayleigh fading channels. We formulate the problem into a manageable optimization problem, and get a closed form scheduler under time invariant channels and time varying channel with finite channel states. For Rayleigh fading channels, we focus on the relay node operation and propose heuristic power allocation algorithm resemblant to water filling algorithm. By simulations, we find that even if the channel fluctuates randomly, the average time span is minimized when the relay node transmits/schedules network coded data as much as possible.
\end{abstract}


\section{Introduction}

Network coding attracted the attention in wireless networks \cite{fragouli2006network}. The network coding benefit is easily represented by reducing the number of transmissions. For example, in two-way relay networks \cite{rankov2006achievable} of Figure \ref{F:twoway}, the traditional method requires four slots to exchange data, whereas the network coded relaying requires three slots \cite{larsson2005coded}. In \cite{oechtering2008broadcast}, moreover, two-slot transmission strategy is studied. In the example, the relay node should have data from both source nodes to execute network coding. This means the source nodes must transmit
their data before the relay node does. Consequently, scheduling can significantly impact the performance of network coding.

The packet scheduling in multiuser communications was originally motivated by \cite{Knopp1995}, where only one user that has the best channel condition should transmit to achieve the capacity. Since then, many scheduling algorithms have been proposed \cite{shakkottai2003cross}. In \cite{scheuermann2007near}, the authors proposed a cross layer scheme that integrates per-hop packet scheduling, network coding and congestion control. To maximize
coding opportunities, they investigated the scheduling of transmissions. However, they considered only time invariant case, where all links have the identical channel gain. In \cite{sagduyu2008throughput}, the authors considered a throughput optimal scheduling problem with random traffic. They showed that both digital and analog network coding outperform the plain routing. In \cite{Chaporkar2007}, the authors pointed that if network coding
and scheduling are designed separately, the expected throughput gain may not be achieved. When it comes to throughput maximization criterion, the buffers of nodes are assumed full. The full buffer assumption is not realistic though it may provide tractableness of analysis.

The network situation can arbitrarily change over time by node mobility, channel fading and the amount of data. Considering these dynamics, throughput may not be well defined. Instead, the time it takes to transfer a bunch of data volume at each source can be practical measurement of the network performance \cite{Pantelidou2009}. In this point of view, we consider the finite amount of initial data traffic by introducing a \emph{time span minimization} criterion \cite{jantti2001transmission}. The details of time span minimization are explained later.

Scheduling in time varying channels is a difficult problem. In \cite{liu2003framework}, the opportunistic scheduling, which exploits time varying nature of the radio environment, is proposed. Scheduling algorithms with consideration of energy-efficiency \cite{lee2008energy}, limited channel information \cite{Chaporkar2009} are investigated.

In this paper, we propose a scheduling scheme incorporating network coding and channel varying information for the two-way relay networks. The provided channel information will constrain the choice of the scheduling policy. If we know the current and future instantaneous channel state information, we can adjust appropriate transmission rates. However, in practical situations, we know only the current channel information and the distribution of the channel fluctuation. Our question is how to design scheduling algorithms incorporating network coding and channel varying information. We consider three channel models, time invariant, time varying with finite states and Rayleigh fading. The first two models may less reflect practical situations than the Rayleigh fading model. However, we can obtain an optimal solution by analysis in the first two cases, which helps us understanding the relation between scheduling and network coding. In Rayleigh fading model, we propose a heuristic algorithm and obtain meaningful
results by simulations.

Throughout the paper, we use the time span as the performance metric for our scheduling algorithm. The time span means the amount of time needed to send all the data in every source node in the network. The problem was first defined in uplink packet scheduling for the cellular CDMA system \cite{jantti2001transmission}. By minimizing
the time span, we can shorten the end-to-end delay while keeping the per-node throughput reasonably high. The rationale behind is to take care of both throughput and delay.

The rest of the paper is organized as follows. In Section 2, we describe the system model. Our scheduling scheme for the minimum time span is proposed in Section 3. In Section 4, we verify the performance of the proposed scheme by simulations. Section 5 concludes the paper.

\section{System Model}

\begin{figure}[tb]
\centerline{\epsfig{figure=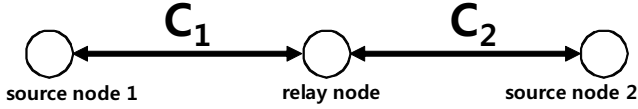,width=3.5in}} \caption{The
two-way relay networks employing network coding at relay node.}
\label{F:twoway}
\end{figure}

Consider the two-way relay networks as shown in Figure \ref{F:twoway}. The source nodes at both ends communicate with each other using the relay node in the middle. The relay node carries out network coding (e.g., eXclusive OR) operations to the received data from two source nodes. $C_{i}$ is the maximum data rate through the link $i$ during a time slot. $C_{i}$ is constant, but randomly varies between slots due to the time varying channel. The channel model between the two nodes is reciprocal. Then, the link is symmetric in that both directions have the same rate. We consider two models for the time varying channel model. The first one is a finite channel state model. Each data rate varies in a set of finite possible rates $\mathbf{S} = (s_1\ \ldots\ s_n)$. Without loss of
generality, we assume $0 < s_1 < s_2 < \ldots < s_n$. We represent the channel state as a vector of data rates $(C_1\ C_2)$. In a time slot, the channel state is one of possible combinations of data rates. For example, if $n=2$, then there are four channel states: $\mathbf{S}^{(1)} = (s_1\ s_1)$, $\mathbf{S}^{(2)} = (s_1\ s_2)$,
$\mathbf{S}^{(3)} = (s_2\ s_1)$ and $\mathbf{S}^{(4)} = (s_2\ s_2)$. We denote the probability that the channel state is $\mathbf{S}^{(i)}$ as $p^{(i)}$. In a time slot, if the \emph{assumed} channel state vector (i.e., the rates using a specific level of modulation and coding) is element-wisely less than or equal to the \emph{actual} channel state vector, then transmissions succeed. The second model is the continuous state model of Rayleigh fading. Each channel power gain follows an exponential distribution with a mean value.

Let $B_i$ denote the finite amount of data that source node $i$ wants to send to the other source node. We assume only one node can transmit at any time instant, i.e., TDMA.

\begin{figure*}[tb]
\centering
\mbox{%
    \subfigure[${\mathbf{S}}^{(1)}=(1\ 1)$]{%
        \includegraphics[width=1.75in]{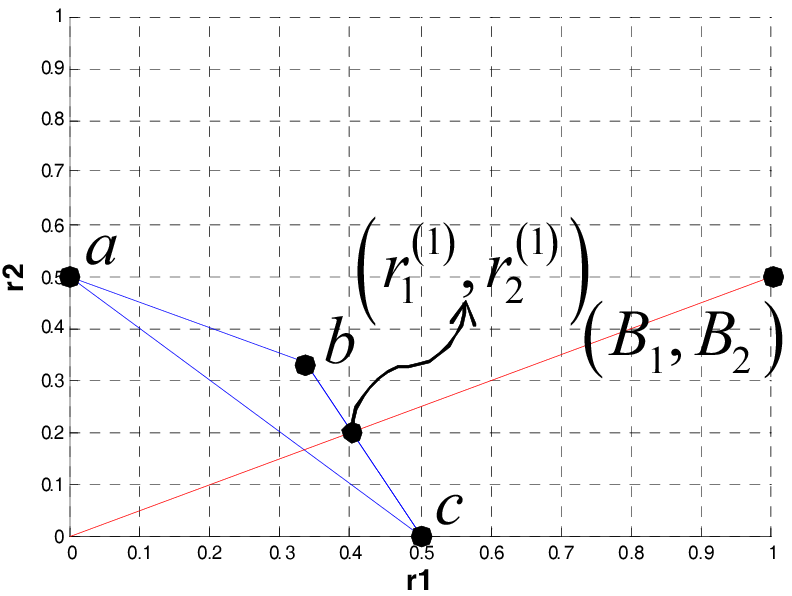}
        \label{F:varying_rate_region1}
    }
    \subfigure[${\mathbf{S}}^{(2)}=(1\ 2)$]{%
        \includegraphics[width=1.75in]{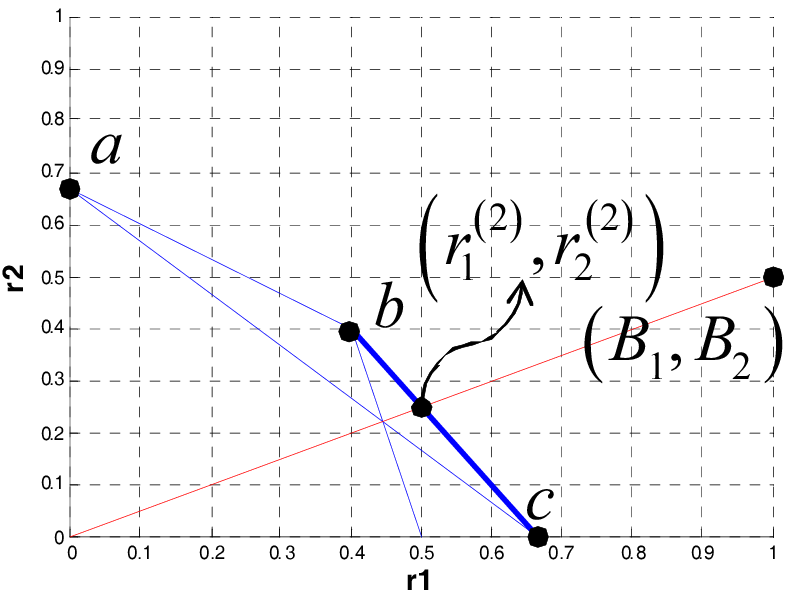}
        \label{F:varying_rate_region2}
    }
    \subfigure[${\mathbf{S}}^{(3)}=(2\ 1)$]{%
        \includegraphics[width=1.75in]{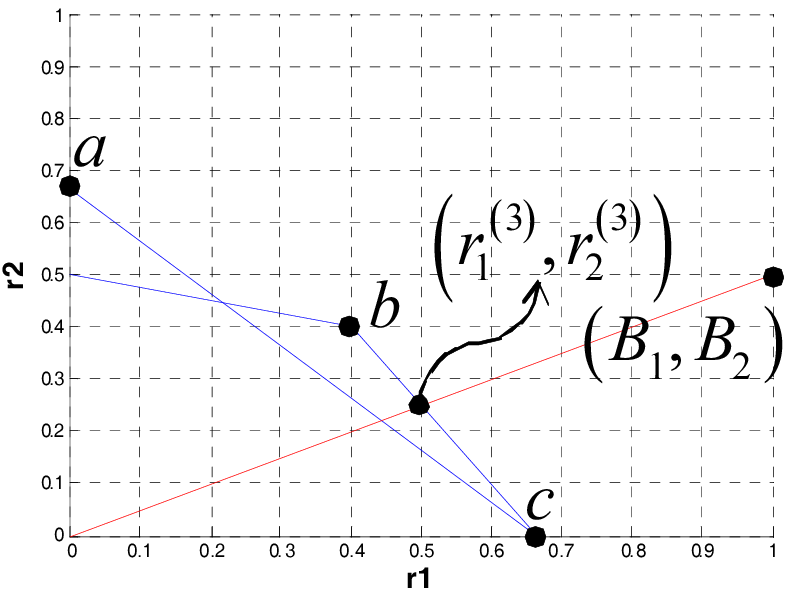}
        \label{F:varying_rate_region3}
    }
    \subfigure[${\mathbf{S}}^{(4)}=(2\ 2)$]{%
        \includegraphics[width=1.75in]{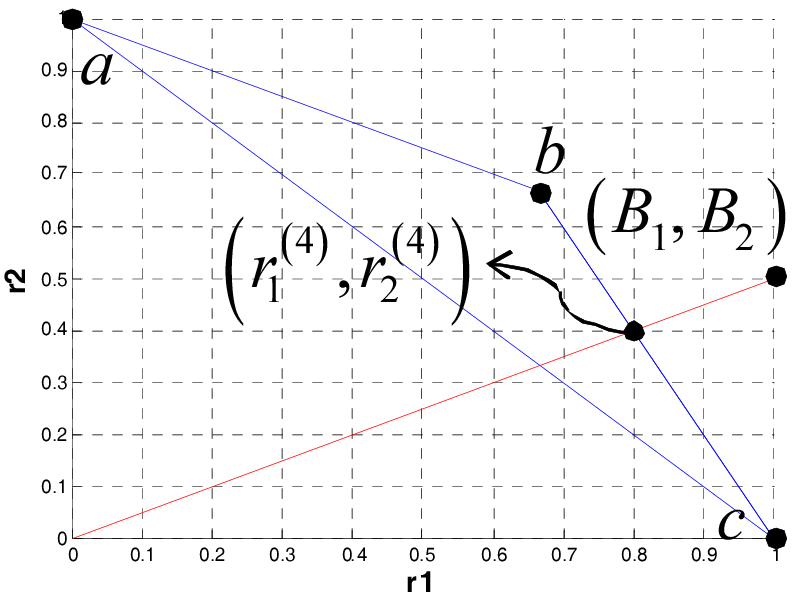}
        \label{F:varying_rate_region4}
    }
} \caption{Four instantaneous rate regions of time varying channels. Channel states are ${\mathbf{S}}^{(1)}=(1\ 1),{\mathbf{S}}^{(2)}=(1\ 2),{\mathbf{S}}^{(3)}=(2\ 1)$ and ${\mathbf{S}}^{(4)}=(2\ 2)$.}\label{F:varying_rate_region}
\end{figure*}

\section{Time Span Minimization}

To propose a scheduling scheme for the minimum time span, we start with the time invariant channels, and extend the result to the case of randomly varying channels.

\subsection{Time Invariant Channels}

It is a special case of the time varying channels; one of $p^{(i)}$'s is one, and the others are zero. To begin with, consider the rate region that depicts the amount of \emph{transmittable} data in a single time slot (see Figure \ref{F:varying_rate_region1}). The rate region \cite{Liu2008} is a convex set and the network coding is
adopted by the relay node. In the figure, the corner point $b=(r_{nc},r_{nc})$ represents the data rate of the (network coded) three-mode transmission. In the first fraction of a time slot, a source node sends data to the relay node. In the second fraction, the other source transmits. In the last fraction, the relay node broadcasts the network coded data to both ends. Note that $r_{nc}$ is the maximum achievable rate by network coding to exchange data between source nodes. The network coded data rates of both end nodes are equal. This is because the network coded data rates are symmetric and bounded by the minimum of maximum transmission rates of links. Note that $r_j$ is the maximum achievable rate by one-directional forwarding from source node $j$
to the other source node. Two edge points $a=(0,r_2)$ and $c=(r_1,0)$ represent one-directional forwarding. The line segment between the corner point ($b$) and an edge point ($a$ or $c$) denotes the time multiplexing of network coding and one-directional forwarding. From Theorems 3.1 and 3.2 of \cite{Liu2008}, the maximum rates of one-directional forwarding and network coding are:
\begin{align}\label{E:netcodrate}
 r_1  = r_2  = \left( {\frac{1}{{C_1 }} + \frac{1}{{C_2 }}} \right)^{ - 1} ,\\r_{nc}  = \left( {\frac{1}{{C_1}} + \frac{1}{{C_2 }}} + \frac{1}{{\min{\left(C_1,C_2\right)} }}\right)^{ - 1}.
\end{align}

To minimize the time span, we formulate the minimum time span into the following linear programming problem:
\begin{align}\label{E:tsm_invariant}
 &\min \theta_{1} + \theta_{2}  + \theta_{3} \nonumber\\
 &{\rm{\ s.t.\ }}\theta_{1} \left( \begin{array}{l}
 r_1  \\
 0 \\
 \end{array} \right) + \theta_{2} \left( \begin{array}{l}
 0 \\
 r_2  \\
 \end{array} \right) + \theta_{3} \left( \begin{array}{l}
 r_{nc}  \\
 r_{nc}  \\
 \end{array} \right) = \left( \begin{array}{l}
 B_1  \\
 B_2  \\
 \end{array} \right) \nonumber\\
 &{\rm{\ \ \ \ \  }}0 \le \theta_{1}  \le \frac{{B_1 }}{{r_1 }},0 \le \theta_{2}  \le \frac{{B_2 }}{{r_2 }},0 \le \theta_{3}  \le \frac{{\min \left( {B_1 ,B_2 } \right)}}{{r_{nc} }}
 \nonumber\\
 &{\rm{\ \ \ \ \  }}\theta_1, \theta_2, \theta_3 \ge 0,
\end{align}
where $\theta_{1}$ denotes the transmission time of source node $1$ to source node $2$ by one-directional forwarding, and $\theta_{2}$ corresponds to that of source node $2$. The time for network coded transmission is denoted by $\theta_{3}$. The maximum network coded data size is $\min(B_1, B_2)$ because the relay node codes the same amount of received data from both source nodes. The objective is to minimize the total transmission time to send the target amount of data $(B_1, B_2)$.

%\vskip 10pt
\noindent {\bf Proposition 1.} {\it To achieve the minimum time span, the time for network coding operation should be $\theta_{3}^* = {{\min \left( {B_1 ,B_2 } \right)}}/{{r_{nc} }}$. }

\noindent {\bf Proof.} We can rewrite the first constraint of \eqref{E:tsm_invariant} in terms of $\theta_3$ as follows:
\begin{equation*}
\theta_{1} = \frac{{B_1 - \theta_{3} r_{nc} }}{{r_1 }},\theta_{2} =
\frac{{B_2 - \theta_{3} r_{nc} }}{{r_2 }},\theta_{3}  =
\theta_{3}.\end{equation*} Then objective function of \eqref{E:tsm_invariant} can be rewritten as:
\begin{equation*}
\frac{{B_1 }}{{r_1 }} + \frac{{B_2 }}{{r_2 }} + \theta_{3} \left( {1
- r_{nc} \left( {\frac{1}{{r_1 }} + \frac{1}{{r_2 }}} \right)}
\right).
\end{equation*}
Without loss of generality, we assume $C_1 \geq C_2$ and from \eqref{E:netcodrate}:
\begin{equation*}
 r_{nc} \left( {\frac{1}{{r_1 }} + \frac{1}{{r_2 }}}
\right) = \left( {\frac{1}{{C_1 }} + \frac{2}{{C_2 }}} \right)^{ -
1} \left( {\frac{2}{{C_1 }} + \frac{2}{{C_2 }}} \right) =
\frac{{2C_1  + 2C_2
}}{{2C_1  + C_2 }}.
\end{equation*}
Because $\frac{{2C_1  + 2C_2 }}{{2C_1  + C_2 }} > 1$, it is $\theta_3\left(1 - r_{nc} \left( {\frac{1}{{r_1 }} + \frac{1}{{r_2}}} \right)\right) < 0$.

\noindent To minimize the objective function, $\theta_3$ should be as large as possible. Therefore, $\theta_{3} = {\min \left( {B_1,B_2 }\right)}/{r_{nc}}. \hfill \Box$%\vskip 10pt

The consequence of Proposition 1 is summarized as follows: First, either node 1 or node 2 transmits its data during ${B_1}/{C_1}$ or ${B_2}/{C_2}$, respectively. Then, the relay node transmits the received data with network coding during ${\min \left(B_1,B_2\right)}/{\min\left(C_1, C_2\right)}$ and with the one-directional forwarding for flushing the remaining data during ${\left(B_1 - B_2\right)}/{C_2 }$ or ${\left(B_2 - B_1\right)}/{C_1}$.

Proposition 1 explains what the solution of the time span minimization looks like but a question still remains regarding how to achieve it. We propose a rate region based scheduling to achieve the minimum time span. We draw a line segment across $(0,0)$ and $(B_1, B_2)$ as in Figure \ref{F:varying_rate_region}. The drawn line has the slope ${B_2}/{B_1}$. There is a cross point $(r_1^{(i)}, r_2^{(i)})$ between the rate region and the line. The number $r_j^{(i)}$ denotes the data rate of node $j$ under the channel state $i$. To minimize the time span, each source node should transmit $r_1^{(i)}$ and $r_2^{(i)}$ under the channel state $i$, by time multiplexing of
network coding and one-directional forwarding. Then a time slot is over. We draw another line between $(0,0)$ and the remaining data $(B_1-r_1^{(i)}, B_2-r_2^{(i)})$ and repeat the same procedure. This intuitively explains how the minimum time span scheduling works.

\subsection{Time Varying Channels}

\subsubsection{Finite State Model}

Consider the channels vary randomly with channel states. For example, if the $n=2$, we can draw four instantaneous rate regions as in Figure \ref{F:varying_rate_region}. Nodes hardly find out the current channel state, but know the probability $p_i$ that the channel state is $\mathbf{S}^{(i)}$. Due to the randomness of the
channel, the \emph{expected} time span will depend on the scheduling policy. If a source node selects the rate region of ${\mathbf{S}}^{(4)}$, the transmission succeeds only when the actual rate region is ${\mathbf{S}}^{(4)}$. However, if the source node selects the rate region of ${\mathbf{S}}^{(1)}$, its transmission succeeds for all channel realizations.

We want to minimize the expected time span. For the purpose, let us consider a probabilistic scheduling policy such that the probability to select $\mathbf{S}^{(i)}$ is $q_i$. The expected data rate of source node $j$ in ${\mathbf{S}}^{(i)}$ is defined as ${\sum\limits_{k \in A^{\left( i \right)} } {p_k r_j^{\left( i\right)} } }$, where $A^{\left( i \right)}  = \left\{{k|{\mathbf{S}}^{\left( k \right)} \preceq {\mathbf{S}}^{\left( i\right)} } \right\}$ is the index set of transmittable channel states $i$. $E[r_j]$ denotes the expectation of an instantaneous data rate of node $j$. For two vectors $\mathbf{a}$ and $\mathbf{b}$, we define operator $\mathbf{a} \preceq \mathbf{b}$ if $\mathbf{a}$ is element-wisely less than or equal to $\mathbf{b}$. Similar to the time invariant case, the ratio of expected data rates $E[r_2]/E[r_1]$ should be $B_2/B_1$ to minimize the time span. We can derive expected data rate of node $1$ as:
\begin{equation}
E\left[ {r_1 } \right] = E_{q_i } \left[ {E\left[ {r_1
|{\mathbf{S}}^{\left( i \right)} } \right]} \right] = \sum\limits_{i
= 1}^{n^2} {\left( {\sum\limits_{j \in A^{\left( i \right)} }^{n^2} {p_j
r_1^{\left( i \right)} } } \right)q_i }.
\end{equation}

We assume finite amount of data that source nodes want to send. In this case, the time span minimization problem can be solved by solving expected data rate maximization problem. To this end, we consider the following optimization problem:
\begin{equation}\label{E:tsm_lp}
\begin{array}{l}
 \mathop {\max }\limits_{q_1 , \ldots ,q_n} \sum\limits_{i = 1}^{n^2}{\alpha_i q_i}  \\
 {\rm{subject\ to\ \ }}\sum\limits_{i = 1}^{n^2} {q_i }  = 1 \\
 {\rm{\ \ \ \ \ \ \ \ \ \ \ \ \ \ \ }}q_i  \ge 0,i=1, 2,..., n.,
 \end{array}
\end{equation}
where $\alpha_i = {\sum\limits_{k \in A^{\left( i \right)} } {p_k\left(r_1^{\left( i \right)} + r_2^{\left( i \right)}\right) } } = \left({1 + \frac{{B_2 }}{{B_1 }}} \right){\sum\limits_{k \in A^{\left( i\right)} } {p_k r_1^{\left( i \right)} } }$ is the sum of expected data rates of source nodes 1 and 2 when channel state $i$ is
selected.

In \eqref{E:tsm_lp}, the linear programming is solved by assigning 1 to $q_i$ that has the largest coefficient:
\begin{equation}\label{E:optimal_q}
q_i^* = 1, {\rm{\ for\ }}i = \arg\max\alpha_i.
\end{equation}

Each node selects the channel state using $q_i^*$ and decides its transmission rate. Even if the channel fluctuates randomly, each source transmits/schedules its data as if the channel were fixed. Consequently, the time for network coded transmission can be find by Proposition 1. The above scheduling configuration is performed at
the beginning of each time slot to decide transmission order and data rate. The node that decides scheduling policy (the relay node is favorable) must know initial data size of the other nodes. Due to the nature of two-way relay networks, the data size information of the other source nodes is readily available. We can apply the following protocol to share data size information. First, relay node requests source nodes to inform their data sizes. Source nodes send data size information to the relay node, respectively. After gathering data size information, the relay node computes the transmission rates of sources by \eqref{E:tsm_lp}. It broadcasts these scheduling information to sources.

\subsubsection{Rayleigh Fading Model}

\begin{figure}
\centerline{\epsfig{figure=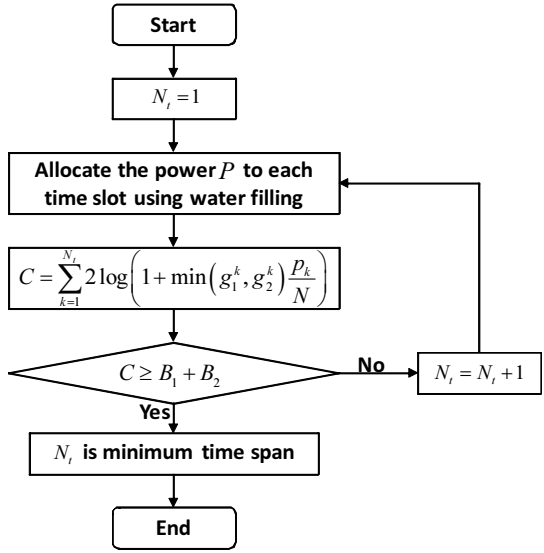,width=2.8in}}
\caption{Time span minimization with noncausal channel knowledge.}
\label{F:wf_noncausal}
\end{figure}

\begin{figure}
\centerline{\epsfig{figure=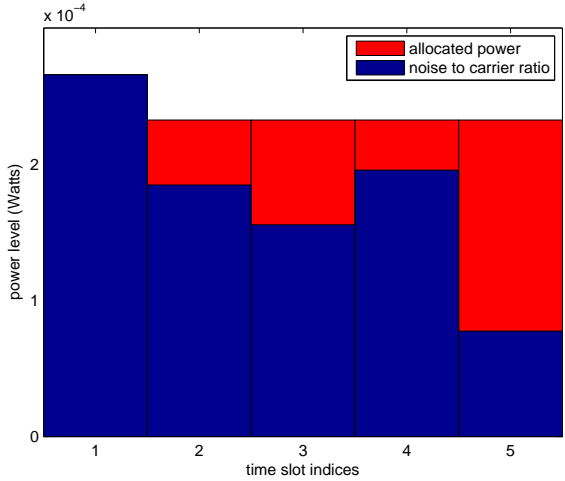,width=3.5in}}
\caption{Water filling power allocation with noncausal channel
knowledge.} \label{F:wf_noncausal_-5dbm}
\end{figure}

We assume now that channels experience the Rayleigh fading. The channel power gain $g_i$ varies between slots following the exponential distribution with a known parameter $\mu$. In this case, the approach in the previous
subsections is inapplicable and analytical results may not be found. We introduce a total power constraint $P$ (W), the additive white Gaussian noise power $I$ (W/Hz) and focus on the relay node operation. We assume that the relay node has received all the data from sources. According to Proposition 1, to minimize time span, the
relay node should have at least $\min \left( {B_1 ,B_2 } \right)$ data from each of sources. Then the problem is reduced to the power allocation problem over slots. The water filling is known to be optimal for allocating powers to orthogonal multi-channels in throughput maximization. We have the channel information \emph{causally} if the past and current channel information are available. On the other hand, we have the channel information
\emph{noncausally} if the future channel information is known. In the noncausal channel information case, we can regard each time slot as a orthogonal subchannel block. If the number of time slots is fixed, we can solve the throughput maximization problem by the water filling power allocation.

Given the fixed amount of data, the time span is minimized by throughput maximization at the relay node. If the amount of received data from both sources is equivalent ($B_1 = B_2$), the relay node can always transmit network coded data. We consider the following optimization problem for relay node:
\begin{equation}\label{E:wf_tsm}
\begin{array}{l}
 \min N_t  \\
 {\rm{subject\ to\ \ }}\sum\limits_{k = 1}^{N_t } {2\log \left( {1 + \min \left( {g_1^k ,g_2^k } \right)\frac{{p_k }}{I}} \right)}  \ge B_1 + B_2 \\
 {\rm{\ \ \ \ \ \ \ \ \ \ \ \ \ \ \ }}\sum\limits_{k = 1}^{N_t } {p_k }  \le P \\
 {\rm{\ \ \ \ \ \ \ \ \ \ \ \ \ \ \ }}p_k  \ge 0,\forall k, \\
 \end{array}
\end{equation}
where, $N_t$ is the number of required time slots to transmit the given data and $g_i^k$ is channel power gain for link $i$ at time slot $k$. The total throughput over the time slots should be greater than or equal to the received data size to flush out the data. We propose to solve the problem in the following way. The procedure is
also depicted in Figure \ref{F:wf_noncausal}. First, assuming $N_t=1$, allocate the power budget $P$ to the first time slot. If the total throughput over the slot is greater than or equal to the remaining data size, a feasible solution is founded. If not, increase one time slot $N_t=2$, and, apply the water filling allocation again with the power budget $P$. We continue this procedure until flushing out the remaining data. Figure \ref{F:wf_noncausal_-5dbm} is an example of the power allocation. The total power constraint is -5 dBm and the data size is 30 MBytes. Five slots are consumed to transmit all data.

\begin{figure}
\centerline{\epsfig{figure=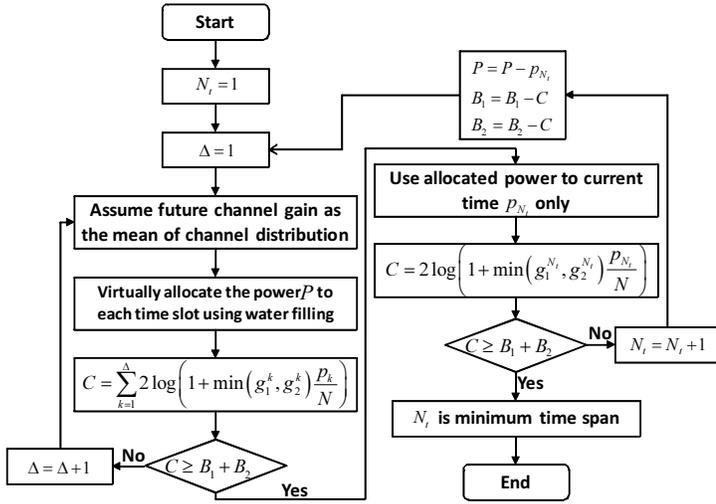,width=3.75in}}
\caption{Time span minimization with causal channel knowledge.}
\label{F:wf_causal}
\end{figure}

\begin{figure}
\centerline{\epsfig{figure=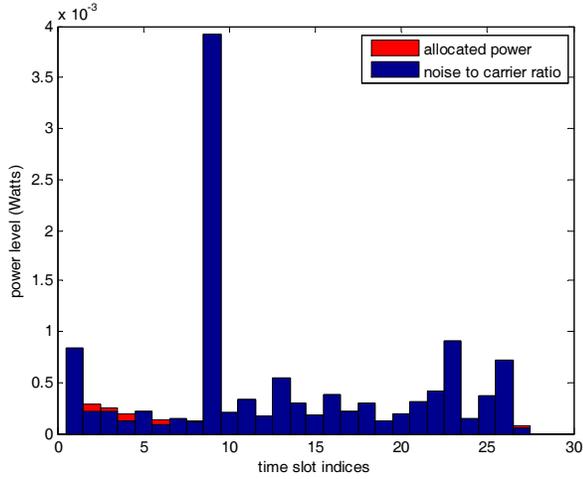,width=3.5in}}
\caption{Heuristic power allocation with causal channel knowledge.}
\label{F:wf_causal_-6dbm}
\end{figure}

In practice, the above algorithm may not be applicable because the future channel knowledge cannot be obtained. On the other hand, if we know the channel distribution and its mean value, the following heuristic approach can be used instead. The algorithm is described in Figure \ref{F:wf_causal}. First, assuming $N_t=1$ and a virtual
time slot $\Delta = 1$, then virtually apply the water filling allocation with power budget $P$. If the total throughput over the time slots is greater than or equal to the remaining data size, a feasible solution is founded. If not, increase one virtual time slot $\Delta + 1$, assume that the future channel is the same as the mean value of the channel distribution and virtually apply the water filling allocation again. When flushing out the remaining data, the power $p_{N_t}$ allocated the current time slot, is actually used. Recalculate the power budget and the remaining data size. We continue this procedure until the relay node has no remaining data. Figure \ref{F:wf_causal_-6dbm} is an example of the heuristic power allocation. Total power constraint is -6 dBm and data size is 15 MBytes. Total 26 time slots are consumed to transmit the given data. The water level is not flat because actual channel information is not given. However, the power is allocated to relatively high channel quality slots. This means the heuristic power allocation works like the water filling.

The heuristic algorithm can be combined with one-directional forwarding. Before the power allocation step, the relay node compares the throughput of three cases, i.e., network coding and two one-directional forwarding transmissions, and selects the transmission scheme shown the largest throughput at each time slot. The other type of incorporation is also possible. For example, we can put high priority on the network coding. If network coding is possible at the current time slot, it is selected even though one-directional forwarding shows higher throughput. In this way, a heuristic algorithm combined with one-directional forwarding can be applied for the case that the data sizes from two sources are different. The performance of this algorithm is verified in following section.

\section{Simulation Results}

\begin{figure}[b]
\centerline{\epsfig{figure=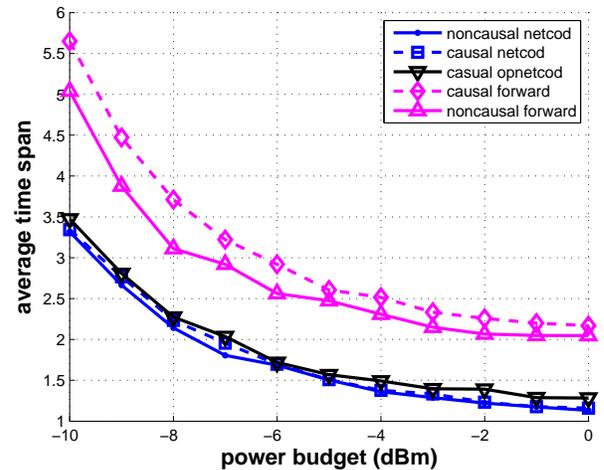,width=3.5in}} \caption{The
average time span as a function of the power budget.}
\label{F:wf_tsm}
\end{figure}

We simulated our heuristic power allocation algorithms under the Rayleigh fading channel. We assume that channel power gains follow the exponential distribution with mean 1. The amounts of data to transmit are $B_1=B_2=8.5$ MBytes. Channel bandwidth is 100 MHz. The noise power is fixed to $10^{-12}$ W/Hz. Figure \ref{F:wf_tsm} shows
the average time span as a function of the power budget. For comparison, noncausal and causal one-directional forwarding cases are plotted together. To execute one-directional forwarding, the relay should decide to whom it should transmit, i.e., the relay node should select one of the two links. The relay node chooses the highest
quality channel between the two channels. With noncausal channel knowledge, the relay node can select future channels in advance. After that, the relay node allocates power using water filling. In the causal case, the relay node can select only one of current channels and allocates power with mean valued future channel gain. As the power budget increases, the average time span decreases for all the cases. Network coding outperforms
one-directional forwarding in all cases considered. The performance gap becomes bigger under a small power budget. With -10 dBm power budget, the gap between the network coding with causal channel information and one-direction forwarding with causal channel information is about 2.2 slots. However, with 0 dBm power budget, the gap is about 1. This result shows that network coding improves the power efficiency. The network coding with the causal channel information, i.e., only executing network coding at the relay node, shows almost the same performance with the network coding with the noncausal channel information as the power budget increases. However, the opportunistic network coding with causal channel information, i.e., selecting proper one between network coding and one-direction forwarding at the current time slot, shows worse performance than the network coding with causal channel information. This result means that network coding opportunity should be maximized to minimize the time span. The imbalance between two buffers can be occurred by the opportunistic
network coding and it reduces network coding opportunity.

\begin{figure}
\centerline{\epsfig{figure=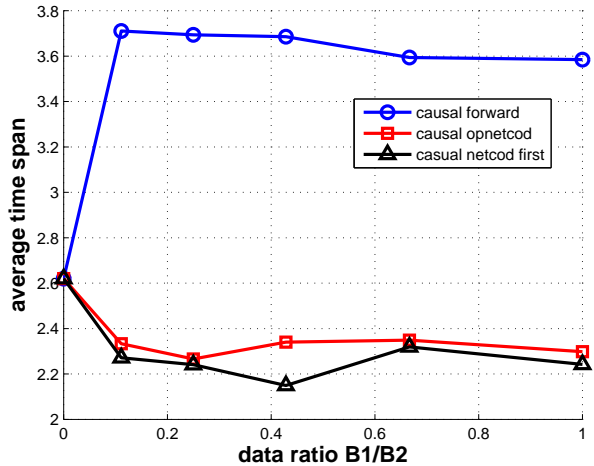,width=3.5in}}
\caption{The average time span as a function of data ratio
($B_1/B_2$).} \label{F:wf_tsm_data_ratio}
\end{figure}

To further investigate the relation between time span minimization and network coding, we simulated opportunistic network coding, network coding first scheme (high priority on the network coding) and one-directional forwarding with various data ratio ($B_1/B_2$). $B_1 + B_2$ is fixed to 17 MBytes. Due to the imbalance between buffers, one-directional forwarding should be used to flush the remaining data. Figure \ref{F:wf_tsm_data_ratio} shows the average time span as the data ratio varies. When the data ratio is zero, network coding cannot be applied and one-directional forwarding is the only option. When the data ratio becomes positive, one-directional forwarding (solid line with circle marker) shows poor performance because the relay node should transmit the both direction. In all cases, the network coding first scheme (solid line with triangle marker) outperforms the opportunistic network coding (solid line with square marker). The results mean that maximizing the network coding opportunity is a better strategy than maximizing the throughput of the current slot.

\section{Concluding Remarks}

We proposed a scheduling scheme incorporating network
coding and channel varying information for the network coded
two-way relay networks. In time varying channels with finite states,
to minimize the expected time span, the proposed scheduler operates
as if channels stay in the specific channel state. Even if the
channel experiences a constant mean valued Rayleigh random fading,
the average time span can be minimized when the relay node
transmits/schedules network coded data as much as possible. A possible research direction is to find
an optimal network coding scheduling under the Rayleigh fading
channel and to extend the results, considering other network coding
strategies (e.g., physical-layer network coding).

\section*{Acknowledgment}

This research was supported by Basic Science Research Program
through the National Research Foundation of Korea (NRF) funded by the
Ministry of Education, Science and Technology (2009-0088483).

\end{document}